\documentclass[review]{elsarticle}

\usepackage{lineno,hyperref,bm,float,amsfonts}
\usepackage{xcolor,graphicx}
\hypersetup{
    colorlinks,
    linkcolor={red!50!black},
    citecolor={blue!50!black},
    urlcolor={blue!80!black}
}
\mathchardef\mhyphen="2D 

\hypersetup{bookmarksopen=true}
\begin{document}

\begin{frontmatter}
\title{Variability response functions for statically determinate beams with arbitrary nonlinear constitutive laws}

\author[a1,a11]{Amir Kazemi}
\ead{amir.kazemi@outlook.com, kazemi.a7@gmail.com}

\author[a2]{Javad Payandehpeyman\corref{cor2}}
\ead{j.payandeh@hut.ac.ir}
\cortext[cor2]{Corresponding author}

\address[a1]{School of Civil Engg., Iran University of Science and Technology, Tehran 16846, Iran.}
\address[a11]{The State Plan and Budget Organization, Tehran 11499, Iran.}
\address[a2]{Department of Robotics Engg., Hamedan University of Technology, Hamedan 65155, Iran.}

\begin{abstract}
The variability response function (VRF) is generalized to statically determinate Euler Bernoulli beams with arbitrary stress-strain laws following Cauchy elastic behavior. The VRF is a Green's function that maps the spectral density function (SDF) of a statistically homogeneous random field describing the correlation structure of input uncertainty to the variance of a response quantity. The appeal of such Green's function is that the variance can be determined for any correlation structure by a trivial computation of a convolution integral. The method introduced in this work derives VRFs in closed form for arbitrary nonlinear Cauchy-elastic constitutive laws and is demonstrated through three examples. It is shown why and how higher order spectra of the random field affect the response variance for nonlinear constitutive laws. In the general sense, the VRF for a statically determinate beam is found to be a matrix kernel whose inner product by a matrix of higher order SDFs and statistical moments is integrated to give the response variance. The resulting VRF matrix is unique regardless of the random field's marginal probability density function (PDF) and SDFs. 
\end{abstract}


\end{frontmatter}

\newpage
\section{Introduction}
The concept of the variability response function (VRF) was introduced in the late 1980s \cite{Shinozuka1987} and has developed extensively since then. The VRF is a means to systematically derive the spectral effects of uncertain system parameters modeled by homogeneous random fields on the response of structures. The VRF is independent from the marginal probability distribution function (PDF) and the spectral density function (SDF) of the random fields.  Using VRFs for a response quantity, one performs the sensitivity of analysis of the system response easily for random fields with different SDFs.

Exact VRFs of displacement response were derived in  \cite{BUCHER1988, DEODATIS1989} for statically determinate beams with linear elastic material. In \cite{Teferra2012}, VRFs were derived for statically determinate beams with power constitutive laws. The concept of the VRF was adapted in \cite{Arwade2011} to measure the variability of upscaled material properties of stochastic volume elements, and to derive VRFs for the effective flexibility of statically determinate beams. 

For statically indeterminate structures, exact VRFs  have not been derived, yet Taylor expansion techniques were used in \cite{Graham1998,Graham2001,Noh2006,Wall1994} for the displacement response of structures whose uncertainty is given by two-dimensional random fields.
Also, the fast Monte Carlo methodology proposed in \cite{SHINOZUKA1988b} was developed in \cite{Papadopoulos2006a,Papadopoulos2006} to estimate the VRF efficiently.
The method was later applied to general linear finite element systems, including dynamic problems \cite{Papadopoulos2012,Papadopoulos2014,giovanis2015adaptive,papadopoulos2015transient} and its ansatz (i.e. the independence of the VRF from the mariginal PDF and SDF of the stochastic field) was examined through the Generalized VRF methodology introduced in \cite{Miranda2012} addressing the static indeterminacy of structures.
The methodology was employed to estimate the VRF for effective flexibility of statically indeterminate beams in \cite{Teferra2012a}, statically indeterminate beams with power constitutive laws in \cite{Teferra2012}, and two-dimensional structures in \cite{Teferra2014}.

The unconditional existence of the VRF nevertheless has neither been proved nor disproved formally under general material nonlinearity. The Generalized VRF methodology, when applied to nonlinear constitutive laws, requires knowing the specific higher order spectral functions affecting response variability. Identifying these higher order terms requires knowing the VRF solution of a statically determinate structure with the same constitutive law.

The derivation presented in this work shows that VRFs can be calculated for statically determinate beams having constitutive laws of arbitrary functional form. The VRFs obtained through this method are a generalization of the classical VRF. By a polynomial interpolation of the beam's curvature in terms of the nominal resisting bending moment, response variance can be expressed as the inner product of a VRF matrix by a matrix of higher order SDFs and statistical moments of the random field describing the resisting bending moment uncertainty. The new formulation results in the same formulas for the VRFs of a linear and square root constitutive law, as well as the same coefficients of higher-order spectral functions \cite{Teferra2012}. Moreover, in a numerical example, the response variance of a stochastic cantilever beam having a bilinear constitutive law is derived using this new approach. Trivial deviation of the results from Monte Carlo (MC) simulations shows that whenever an accurate polynomial interpolation is used to model the curvature in terms of the resisting bending moment, the variance can be calculated by the VRFs precisely.

\section{The response of stochastic beams}
Suppose that the section modulus and constitutive law of a transversely loaded statically determinate Euler-Bernoulli beam vary randomly along the beam's length as
\begin{equation}
\frac{1}{\sigma(x,\varepsilon)S(x)}\equiv
\frac{1+f(x)}{\sigma_0(\varepsilon)S_0}
\end{equation}
where $S_0$ and $\sigma_0(\varepsilon)$ denote the nominal section modulus and constitutive law respectively, and $f(x)$ is a zero-mean, statistically homogeneous random field bounded as $f(x)>-1$  which embeds the uncertainty in material and cross-sectional geometry.

Using the kinematic relationship, which states that
\begin{equation}
y=\rho \varepsilon
\end{equation}
where $y$ is the vertical coordinate from the neutral axis and $\rho$ is the curvature radius, the resulted maximum strain ($\epsilon$) along the beam, as an uncertain structural response (or output) quantity, is the random field satisfying
\begin{equation}
|M(x)|=
\frac{3S(x)}{\epsilon^2(x)}
\int_0^{\epsilon(x)}
\sigma(x,\varepsilon) \varepsilon d\varepsilon
\end{equation}
where $|M(x)|$ is the absolute value of static moment at section $x$. Eqs. (1) and (3) give
\begin{equation}
|M(x)|=\frac{1}{1+f(x)} \left(
\frac{3S_0}{\epsilon^2(x)}
\int_0^{\epsilon(x)}
\sigma_0(\varepsilon) \varepsilon d\varepsilon \right).
\end{equation}
Let
\begin{equation}
m_\epsilon(\epsilon)\equiv
\frac{3S_0}{\epsilon^2}
\int_0^{\epsilon}
\sigma_0(\varepsilon) \varepsilon d\varepsilon.
\end{equation}
where $m_\epsilon(\epsilon)$ is the nominal resisting bending moment corresponding to the maximum strain $\epsilon$. The nominal resisting bending moment in terms of curvature ($k=2\epsilon/h$ with $h$ as the section's height) takes the following form:
\begin{equation}
m(k)\equiv
\frac{3S_0}{k^2}
\int_0^{k}
\sigma_0(\xi h/2) \xi d\xi.
\end{equation}
Employing the definition of nominal resisting bending moment, one concludes from Eq. (4) that
\begin{equation}
m(k(x))=|M(x)|(1+f(x))
\end{equation}
The asymptotic behavior of Eq. (7) is in accordance with that of the initial definition in Eq. (1):
As $f(x)\rightarrow + \infty$, it requires  $k(x)\rightarrow + \infty$ (the infinite flexibility case);
contrariwise, when  $f(x)\rightarrow -1$, it makes  $k(x)\rightarrow 0$ (the infinite rigidity case).
Trivially, $k(x)=0$ for $M(x)=0$.
The reader is cautioned that $f(x)$ must posses an upper-bound so that the resisting bending moment acquire meaningful realizations and the MC simulation becomes feasible.
Therefore, to have a well-posed problem, distributions like the lognormal should be applied to $f(x)$ carefully \cite[p.~9]{teferra2012developments}.

Calculation of $k(x)$ is required to find the second derivative of the beam's deflection and thereof the deflection itself. This is realized by finding the inverse of $m(\cdot)$ using Eqs. (6-7):
\begin{equation}
k(x)=m^{-1}\left(  \vert M(x) \vert\left(1+f(x)\right)\right).
\end{equation}
If $m(\cdot)$ is one-to-one, it is invertible as well. Therefore, the next step is to investigate whether $m(\cdot)$ is increasing or not, that is to say:
\begin{equation}
m'(k)=3S\left(\frac{-2}{k^3}\int_{0}^{k}\sigma\left(\frac{h\xi}{2}\right)\xi d\xi+\frac{1}{k^2}\sigma\left(\frac{hk}{2}\right)k\right)> 0
\end{equation}
which, using Eq. (5), yields
\begin{equation}
\int_{0}^{ \epsilon}\sigma(\varepsilon)\varepsilon d\varepsilon<\frac{1}{2}\sigma( \epsilon) \epsilon^2.
\end{equation}
This inequality holds for almost every constitutive law. As shown in the schematic stress-strain curve of Fig. (1), the left hand side of the inequality is the moment of the dotted area with respect to the stress axis, while the right hand side is that of the total shaded area. 
\\
\begin{figure}[!hb]
\centering
\includegraphics[scale=0.4]{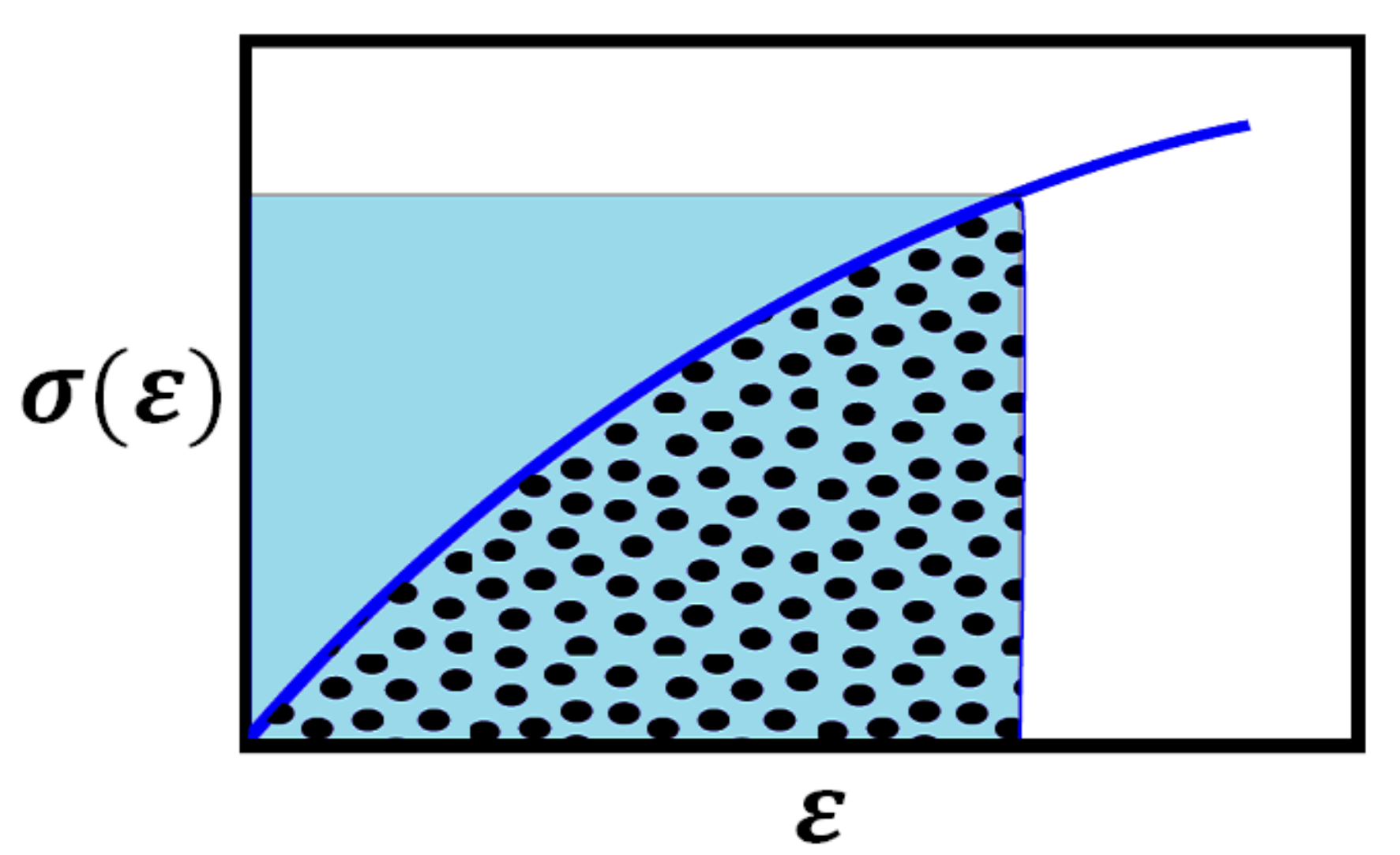}
\caption{Schematic stress-strain curve to show the invertibility of the resisting bending moment in terms of curvature.}
\end{figure}

The curvature or $k(x)$ in Eq. (8) can be approximated by the polynomial interpolation of $m^{-1}(\cdot)$. According to Weierstrass approximation theorem, any continuous function like $m^{-1}(\cdot)$ defined over a closed interval is uniformly approximated by a polynomial as accurately as desired. Without loss of generality, this closed interval in mathematical texts is supposed as $[0,1]$ or $[-1,1]$ to which arbitrary intervals are easily mapped \cite[p.~509]{estep2002practical}. The reader can choose among various polynomial interpolation forms to model curvature in terms of resisting bending moment as a polynomial function.

In this paper,  the monomial form of polynomial interpolation is employed to approximate $m^{-1}(\cdot)$, where the polynomial coefficients are calculated through an explicit formulation constructed from the Vandermonde matrix \cite[~ p. 2]{phillips2003}. The monomial form of polynomial interpolation finds the unique polynomial of $N^{th}$-degree crossing $N+1$ points such that the curvature is expressed as
\begin{equation}
 k(x)=\sum_{i=0}^{N}\lambda_i \left(\vert M(x) \vert (1+f(x))\right)^i
\end{equation}
to fit $\bm{ k}= \left[  k_1,  k_2, ...,  k_{N+1} \right]^T $ and $\bm{m}=[m( k_1), m( k_2), ..., m( k_{N+1})]^T$. The coefficients $\lambda_i$ are calculated by solving a linear system of equations as follows:
\begin{equation}
\bm{\lambda}=\bm{\mathcal{V}}^{-1}\bm{ k}
\end{equation}
\mathversion{normal}
with $\bm{\mathcal{V}}$ defined as the square Vandermonde matrix:
\begin{equation}
\bm{\mathcal{V}}=\left[
\begin{array}{lllll}
1 & m( k_1) & m^2( k_1)  & \cdots & m^N( k_1) \\
1 & m( k_2) & m^2( k_2)  & \cdots & m^N( k_2) \\
\vdots & \vdots & \vdots & \ddots & \vdots \\
1 & m( k_{N+1}) & m^2( k_{N+1})  & \cdots & m^N( k_{N+1}) \\
\end{array} \right].
\end{equation}
It is noteworthy that the interpolation form is called monomial because the bases for the interpolating $N^{th}$-degree polynomial are selected as $1$ $,m( k)$ $,m^2( k)$ $,...,$ $m^N( k)$ which are monomials. 

Note that a major concern for the convergence of polynomial interpolation is Runge's phenomenon which is the oscillation at the edges of the fitting interval including equispaced interpolation points. To minimize the effect of this phenomenon in polynomial interpolation, Chebyshev nodes should be used as fitting data \cite[Ch.~13]{trefethen2013}. For a fitting interval of $[0, k_u]$, the abscissas of such nodes are determined by
\begin{equation}
m( k_n)=\frac{1}{2}m( k_u)+\frac{1}{2}m( k_u)\cos\left(\frac{2n-1}{2N+2}\pi\right).
\end{equation}
The ordinates of the Chebyshev nodes are hence:
\begin{equation}
 k_n=m^{-1}\left(\frac{1}{2}m( k_u)+\frac{1}{2}m( k_u)\cos\left(\frac{2n-1}{2N+2}\pi\right)\right)
\end{equation}
which can be estimated by an interpolation within the pairs of $( k,m( k))$ using Eqs. (6) and (14). Noteworthy is the fact that, according to Eq. (8), a valid interpolation requires
\begin{equation}
 k_u\geq m^{-1}(\max(\vert M(x) \vert(1+f(x))).
\end{equation}

As the kinematic relationship in Eq. (2) states, the beam's signed curvature is given by
\begin{equation}
u''(x)=k(x)\textup{sgn}(M(x))
\end{equation}
which means that the signed curvature is positive under positive static moment. Importing Eq. (11) in Eq. (17) yields
\begin{equation}
u''(x)=\sum_{i=0}^{N}\lambda_{i}\left(\vert M(x)\vert (1+f(x))\right)^i\textup{sgn}(M(x))
\end{equation}
which is solved as
\begin{equation}
u(x)=\int_{0}^{x}\sum_{i=0}^{N}\lambda_{i}\left(\vert M(s) \vert (1+f(s))\right)^i\textup{sgn}(M(s))G(x,s)ds
\end{equation}
where $G(x,s)$ is the Green's function for the differential equation in Eq. (18) along with imposed boundary conditions on $u(x)$.

\section{The VRFs}
The response variance, i.e. $\texttt{Var}[u(x)]=\texttt{E}[u^2(x)]-\texttt{E}[u(x)]^2$, can be written as
\begin{eqnarray}
\nonumber \texttt{Var}[u(x)]= \sum_{i=0}^N\sum_{j=0}^N
\int_0^x\int_0^x \lambda_{i}\lambda_{j}  \\
\nonumber \times \vert M(s_1) \vert ^i \vert M(s_2) \vert ^j
\textup{sgn}(M(s_1))\textup{sgn}(M(s_2)) \\
\times G(x,s_1)G(x,s_2)
\left(R^*_{ij}(\tau)-\mu^*_i\mu^*_j\right) ds_1ds_2
\end{eqnarray}
with $\texttt{E}[(1+f(s))^i]=\mu^*_i$ (the $i^{th}$ moment), $\texttt{E}[(1+f(s_1))^i(1+f(s_2))^j]=R^*_{ij}(\tau)$ (the $ij^{th}$ autocorrelation function), and $\tau=s_2-s_1$. It is worth noting the explicit dependence of the response variance on the higher order correlations of $f(x)$ for arbitrarily nonlinear constitutive law. By using Wiener-Khinchin theorem, which states that $R_{ij}(\tau)=\int_{-\infty}^{+\infty}S_{ij}(\kappa)\exp(\texttt{i}\kappa\tau)d\kappa$, Eq. (20) is expressed as
\begin{equation}
\texttt{Var}[u(x)]=\int_{-\infty}^{+\infty}\textbf{VRF}(x,\kappa):\left(\textbf{S}(\kappa)-\delta(\kappa)\textbf{M}\right)d\kappa
\end{equation}
where : denotes the Frobenius inner product,  $\textbf{S}$ and $\textbf{M}$ are the matrices of SDFs and statistical moments of the random field $1+f(x)$ with the following components:
\begin{eqnarray}
\textbf{S}_{ij}(\kappa)\equiv S^*_{ij}(\kappa)=\sum_{p=0}^i\sum_{q=0}^j{{i}\choose{p}}{{j}\choose{q}} S_{pq}(\kappa), \\
\textbf{M}_{ij}\equiv \mu^*_{i}\mu^*_{j}=\sum_{p=0}^i\sum_{q=0}^j{{i}\choose{p}}{{j}\choose{q}}  \mu_p\mu_q,
\end{eqnarray}
where asterisks denotes that the parameter belongs to $1+f(x)$, rather than $f(x)$ (for which no asterisk is used).
The matrix $\textbf{VRF}$ is given by the following vector multiplication:
\begin{equation}
\textbf{VRF}(x,\kappa)=\textbf{V}^\dagger(x,\kappa)\textbf{V}(x,\kappa)
\end{equation}
where $\textbf{V}^\dagger$ is the conjugate transpose of $\textbf{V}$ and
\begin{equation}
\textbf{V}_i(x,\kappa)\equiv\int_0^x\lambda_{i}\vert M(s) \vert^i\textup{sgn}(M(s))G(x,s)\exp(\texttt{i}\kappa s)ds.
\end{equation}
Note that $R^*_{ij}(\tau)-\mu^*_i\mu^*_j$ is zero when $i=0$ and/or $j=0$. Therefore, the sums in Eq. (20) could start from $i=1$ and $j=1$. Besides, a correct interpolation of curvature with respect to nominal resisting bending moment requires $\lambda_0=0$ as a result of $m(k=0)=0$.

\section{Parametric Examples}
The derivations in sections 2 and 3 are based on the definition of random field for the reciprocal of section modulus by stress, i.e. Eq. (1), rather than for the elastic flexibility. Therefore, it is critical to examine whether this assumption is robust and leads to the same VRFs for linear and a class of non-linear constitutive laws as shown in \cite{BUCHER1988,DEODATIS1989,Teferra2012} respectively:
\subsection{Linear constitutive law}
Let the nominal constitutive law be $\sigma(\varepsilon)=E\varepsilon$. As a result of stochastic material and cross-section (i.e. Eq. (1)), resisting bending moment is a random field along the beam as obtained in Eq. (7) with the nominal value given as
\begin{equation}
m( k)=\frac{3S}{k^2}\int_{0}^{ k}E\left(\frac{h\xi}{2}\right)\xi d\xi=EIk
\end{equation}
where $I=bh^3/12$ is the moment of inertia of the cross section. Consider an interpolation of $m^{-1}(\cdot)$ by a polynomial of second degree using a set of three points $(0,0)$, $( k_1,m( k_1))$, and $( k_2,m( k_2))$. Note that two points suffices inasmuch as $m^{-1}(\cdot)$ is linear, yet three points are selected to show that adding points in the interpolation does not alter the results for the linear constitutive law. The Vandermonde matrix according to Eq. (13) becomes:
\begin{equation}
\bm{\mathcal{V}}=\left[
\begin{array}{lll}
1 & 0 & 0 \\
1 & m( k_1) & m^2( k_1) \\
1 & m( k_2) & m^2( k_2) \\
\end{array} \right]=\left[
\begin{array}{lll}
1 & 0 & 0 \\
1 & \alpha k_1 & \alpha^2  k_1^2 \\
1 & \alpha k_2 & \alpha^2  k_2^2 \\
\end{array} \right]
\end{equation}
where $\alpha=EI$. Introducing the inverse of $\bm{\mathcal{V}}$ into Eq. (12) yields:
\begin{equation}
\left[
\begin{array}{c}
\lambda_0 \\ \lambda_1 \\ \lambda_2
\end{array}
\right]
=\frac{1}{\alpha^3( k_1 k^2_2- k_1^2 k_2)}
\left[
\begin{array}{ccc}
\alpha^3( k_1 k^2_2- k_1^2 k_2) & 0 & 0 \\
\alpha^2(- k^2_2+ k_1^2) &
\alpha^2 k^2_2 &
-\alpha^2 k^2_1 \\
-\alpha(- k_2+ k_1) &
-\alpha k_2 &
\alpha k_1
\end{array} \right]
\left[
\begin{array}{c}
0 \\  k_1 \\  k_2
\end{array}
\right]
\end{equation}
where the polynomial coefficients are solved as
\begin{equation}
\left[
\begin{array}{c}
\lambda_0 \\ \lambda_1 \\ \lambda_2
\end{array}
\right]=
\left[
\begin{array}{c}
0 \\ 1/\alpha \\ 0
\end{array}
\right].
\end{equation}
Using $\bm{\lambda}$ in Eqs. (24) and (25) gives
\begin{equation}
\textbf{VRF}(x,\kappa)=
\left(\frac{1}{EI}\right)^2
\int_0^x\int_0^x M(s_1)M(s_2) G(x,s_1) G(x,s_2) \exp(\texttt{i}\kappa\tau)ds_1ds_2
\end{equation}
and
\begin{equation}
\textbf{S}(\kappa)-\delta(\kappa)\textbf{M}=
S^*_{11}(\kappa)-\delta(\kappa)\mu^*_1\mu^*_1 =S_{11}(\kappa)
\end{equation}

Eqs. (45) and (46) are exactly the widely-known VRF and SDF for a linear constitutive law \cite{BUCHER1988, DEODATIS1989}. Note that it can be shown that $\lambda_i=0$ for all $i \neq 1$ when solving Eqs. (27) and (28) by assuming a higher degree polynomial and solving for vector $\bm{ \lambda}$.

\subsection{Square root constitutive law}

Let the constitutive law be $\sigma(\varepsilon)=E\sqrt{\varepsilon}$. Resisting bending moment is a random field along the beam as obtained in Eq. (7) with the nominal value given as
\begin{equation}
m( k)=\frac{3S}{k^2}\int_{0}^{ k}E\left(\frac{h\xi}{2}\right)^{0.5}\xi d\xi=\frac{12EI}{5\sqrt{2h}}\sqrt{ k}.
\end{equation}
As a starting point, assume a forth-degree polynomial for interpolating $m^{-1}(\cdot)$. The nodes for interpolation  have $[0\;  k_1\;  k_2\;  k_3\;  k_4]$ as ordinates and $[m( k_i)]$ as abscissas. Note that employing parametric nodes and thus an arbitrary fitting interval obviates the need to control Eq. (16), because one may assume $ k_4\geq k_u$ without loss of generality. The Vandermonde matrix becomes

\begin{equation}
\bm{\mathcal{V}}=\left[
\begin{array}{lllll}
1 & 0 & 0 & 0 & 0 \\
1 & \beta k_1^{0.5} & \beta^2  k_1 &\beta^3  k_1^{1.5} & \beta^4  k_1^2 \\
1 & \beta k_2^{0.5} & \beta^2  k_2 &\beta^3  k_2^{1.5} & \beta^4  k_2^2 \\
1 & \beta k_3^{0.5} & \beta^2  k_3 &\beta^3  k_3^{1.5} & \beta^4  k_3^2 \\
1 & \beta k_4^{0.5} & \beta^2  k_4 &\beta^3  k_4^{1.5} & \beta^4  k_4^2
\end{array} \right]
\end{equation}
where $\beta=(12EI)/(5\sqrt{2h})$. Introducing Eq. (48) into Eq. (12) gives
\begin{equation}
\left[
\begin{array}{c}
\lambda_0 \\ \lambda_1 \\ \lambda_2 \\ \lambda_3 \\ \lambda_4 \\
\end{array}
\right]=
\left[
\begin{array}{c}
0 \\ 0 \\ 1/\beta^2 \\ 0 \\ 0
\end{array}
\right].
\end{equation}
Using $\bm{\lambda}$ in Eqs. (21-25) yields
\begin{eqnarray}
\nonumber \textbf{VRF}(x,\kappa)=\int_0^x\int_0^x
\left(\frac{50}{b^2h^5E^2} \right)^2
\vert M(s_1) \vert ^2 \vert M(s_2) \vert ^2 \\
\nonumber \times\textup{sgn}(M(s_1))\textup{sgn}(M(s_2))G(x,s_1)\\
\times  G(x,s_2) 
\exp(\texttt{i}\kappa \tau)ds_1 ds_2
\end{eqnarray}
and
\begin{eqnarray}
\nonumber
\textbf{S}(\kappa)-\delta(\kappa)\textbf{M}=
S^*_{22}(\kappa)-\delta(\kappa)\mu^*_2\mu^*_2 \\
 =4S_{11}(\kappa)+4S_{12}(\kappa)+S_{22}(\kappa)-\delta(\kappa)\sigma_f^4.
\end{eqnarray}
which agree with Eqs. (30) and (31) in Ref. \cite{Teferra2012}. Mathematical induction can show that $\lambda_i=0$ for all $i \neq 2$ when assuming a higher degree polynomial and solving for vector $\bm{ \lambda}$.

\section{Numerical Example}
\subsection{Structural specification}

To show the method's efficiency for the estimation of VRFs for arbitrary Cauchy elastic materials, a bilinear constitutive law is examined for the statically determinate beam shown in Fig. (2) with $M=3500$, $q(x)=50$, $L=16$, $b=1$, $h=\sqrt[3]{12}$, and $G(x,s)=x-s$. The nominal constitutive law is
\begin{equation}
\sigma(\varepsilon)=\left\lbrace
\begin{array}{ll}
E_0\varepsilon & \varepsilon\leq 0.002 \\
0.1E_0(\varepsilon+0.018) & \varepsilon > 0.002
\end{array}
\right.
\end{equation}
with $E_0=7\times10^5$. The resisting bending moment is considered as a statistically homogeneous random field as derived in Eq. (7) whose nominal value is obtained by introducing Eq. (37) into Eq. (6).
The monomial form of polynomial interpolation is employed to model $m^{-1}(\cdot)$ as suggested in section 2. The analytically derived $\textbf{VRF}$ is verified by comparing the predicted variance using the VRFs, i.e. Eq. (21), for the vertical displacement at $x=16$ with that computed by brute-force MC simulation for three different random field models of $f(x)$ as discussed below.

\begin{figure}[!hb]
\centering
\includegraphics[scale=0.5]{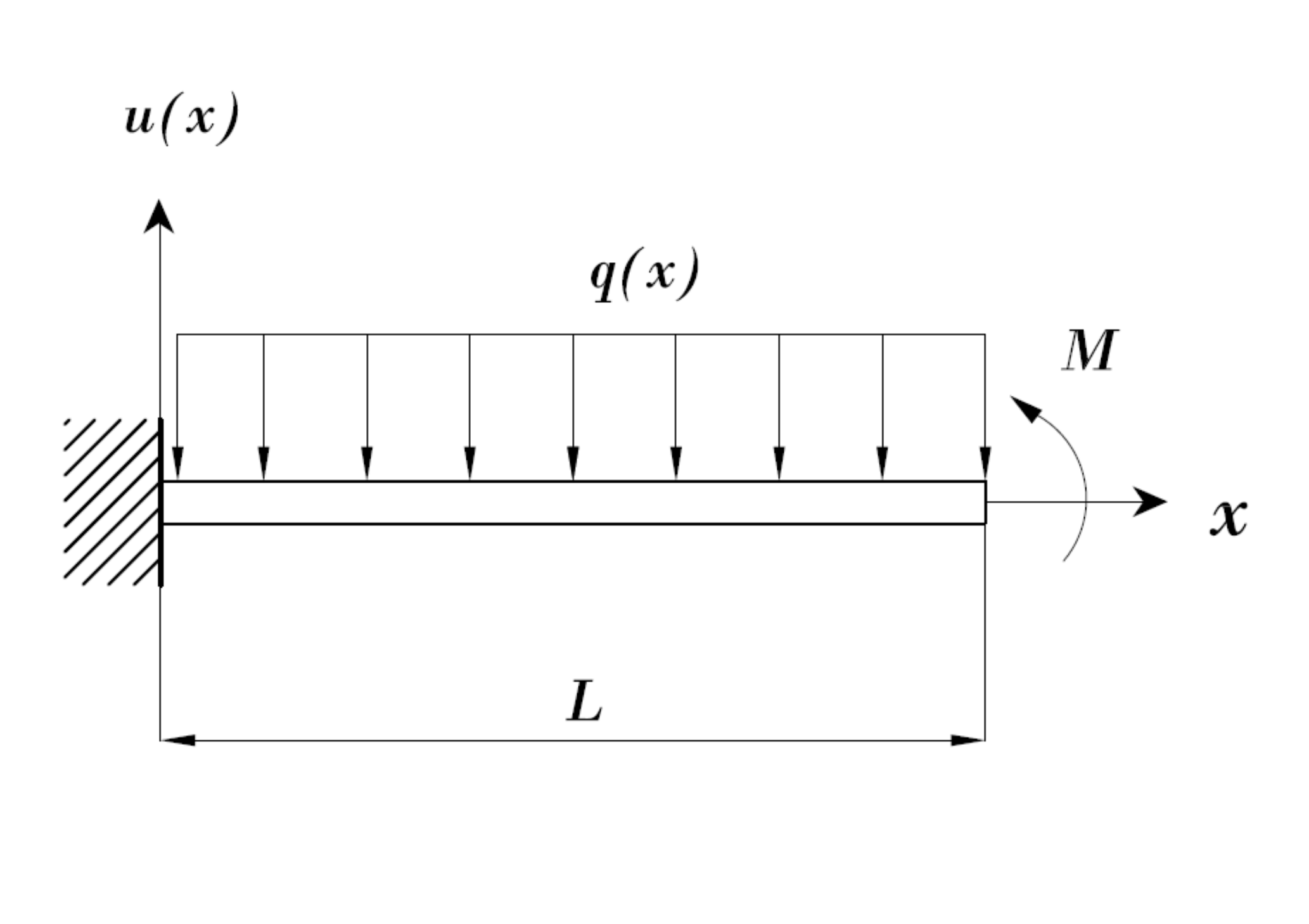}
\caption{Cantilever analysed in the numerical example from Ref. \cite{Teferra2012}}
\end{figure}

\subsection{The associated random field}
The MC simulation employs translation from an underlying U-Beta random field to a target one (an associated field) with a target marginal cumulative distribution function (CDF) $P_f$ \cite{Grigoriu1995,Deo2003a,Deo2003b}. The underlying random field varies sinusoidally with random phase angles $\theta$ uniformly distributed on $[0, 2\pi]$ as follows
\begin{equation}
g(x)=\sqrt{2}\sigma_g\cos{(\kappa_\delta x+\theta)}
\end{equation}
where $\sqrt{2}\sigma_g$ is the amplitude and $\kappa_\delta$ is a certain wave number determining the spectral content of the field. The underlying U-Beta random field has SDF given as $S = \sigma_g/2[(\delta( \kappa + \kappa_\delta) + \delta( \kappa - \kappa_\delta)]$, and the values used in this example are $\kappa_\delta = \pi/2$, and $\sigma_g = 1/\sqrt{2}$.

The associated field is
\begin{equation}
f(x)=P_f^{-1}\circ P_g(g(x))=\mathcal{A}(g(x))
\end{equation}
where $P_g$ denotes the CDF of the underlying field given as
\begin{equation}
P_g(g(x))=
1-\frac{1}{\pi}
\arccos{ \left( \frac{g(x)}{\sqrt{2}\sigma_g} \right) }.
\end{equation}

In this example, the three associated fields considered have uniform (UN), truncated Gaussian (TG), and Lognormal (LN) marginal distributions. The random field $f(x)$ is realized by mapping $g(x)$ as follows: the mapping for UN is given as
\begin{equation}
f(x)=\left(a_u-a_l\right)P_g(g(x))+a_l,
\end{equation}
the mapping for TG is
\begin{equation}
f(x)=\left\lbrace\begin{array}{ll}
a_l & s\Phi^{-1}(P_g(g(x)))+m<a_l  \\
s\Phi^{-1}(P_g(g(x)))+m &
a_l\leq s\Phi^{-1}(P_g(g(x)))+m\leq a_u \\
a_u &  a_u<s\Phi^{-1}(P_g(g(x)))+m
\end{array}
\right. ,
\end{equation}
and the mapping for LN is given as
\begin{equation}
f(x)=\exp{\left(s\Phi^{-1}(P_g(g(x)))+m\right)}+a_l,
\end{equation}
where $a_l,a_u,m$ and $s$ are defined in Table (1). Simulations for $g(x)$ are obtained through the simulation of random variable $\theta$ as given in Eq. (38).

\begin{table}[hb]
\caption{Parameters of the PDFs for $f(x)$ }
\centering
\begin{tabular}{cccccc}
\hline
PDF & $a_l$ & $a_u$ & $m$ & $s$ & $\sigma_f$ \\
\hline
UN & -0.80 & 0.80 & n/a & n/a & 0.46 \\
TG & -0.90 & 0.90 & 0.00 & 1.00 & 0.67 \\
LN &  -0.40 & n/a & -1.03 & 0.47 & 0.20 \\
\hline
\end{tabular}
\end{table}

The SDFs of the associated fields are obtained as follows. Due to the shift invariance of the U-beta random field (i.e. $g(x + 2\pi) = g(x)$) and the one-to-one mapping of the associated field, the autocorrelation function of $f(x)$ is given by
\begin{equation}
R_{ij}(\tau)=\frac{1}{2\pi}\int_0^{2\pi}\mathcal{A}^i(\sqrt{2}\sigma_g\cos{(\theta)})\mathcal{A}^j(\sqrt{2}\sigma_g\cos{(\kappa_\delta\tau+\theta)})d\theta
\end{equation}
which is an even function representable by the following Fourier series:
\begin{equation}
R_{ij}(\tau)=\frac{a_0(i,j)}{2}+\sum_{\eta=1}^\infty a_n(i,j) \cos{(\eta\kappa_\delta\tau)}
\end{equation}
with
\begin{equation}
a_\eta(i,j)=\frac{1}{2\pi^2}\int_0^{2\pi}\int_0^{2\pi}\cos{(\eta\xi)}\mathcal{A}^i(\sqrt{2}\sigma_g\cos{(\theta)})\mathcal{A}^j(\sqrt{2}\sigma_g\cos{(\xi+\theta)})d\theta d\xi.
\end{equation}
Corresponding higher order SDFs, obtained by taking the Fourier transform of Eq. (45), are expressed as
\begin{equation}
S_{ij}(\kappa)=\frac{a_0(i,j)}{2}\delta(\kappa)+\frac{1}{2}\sum_{\eta=1}^\infty a_\eta(i,j)
\left( \delta(\kappa+\eta\kappa_\delta)+\delta(\kappa-\eta\kappa_\delta) \right)
\end{equation}
where $\delta(\cdot)$ is the Dirac's delta function. The statistical moments are also given by
\begin{equation}
\mu_i=\frac{1}{2\pi}\int_0^{2\pi}\mathcal{A}^i(\sqrt{2}\sigma_g\cos{(\theta)})d\theta.
\end{equation}

\subsection{Results and discussion}

As shown in Fig. (3), $m^{-1}(\cdot)$ (the dotted blue line) is fitted by the monomial form of polynomial interpolation (the solid red line) with different degrees using Eqs. (11-16).
The data of $m^{-1}(\cdot)$ is a set of ordered pairs  obtained by interchanging the first and second elements of the pairs $( k,m( k))$ generated by Eq. (6) within the curvature domain $[0,0.1]$.
Such domain, according to Eq. (16), guarantees the validity of the polynomial interpolation for $f(x)$ having the PDF of UN and TG.
Yet for $f(x)$ with the PDF of LN, one should assure that upper tails do not affect the variance significantly. Fig. (4) shows that increasing the truncation value of the LN-based $f(x)$ more than one hardly changes the response variance in the MC simulation.
Therefore, the mentioned curvature domain produces an accurate response for LN truncated as $f(x)\leq 1$.

\begin{figure}[!b]
\centering
\includegraphics[scale=0.61,trim={1.6cm 1cm 1.5cm 1.5cm},clip]{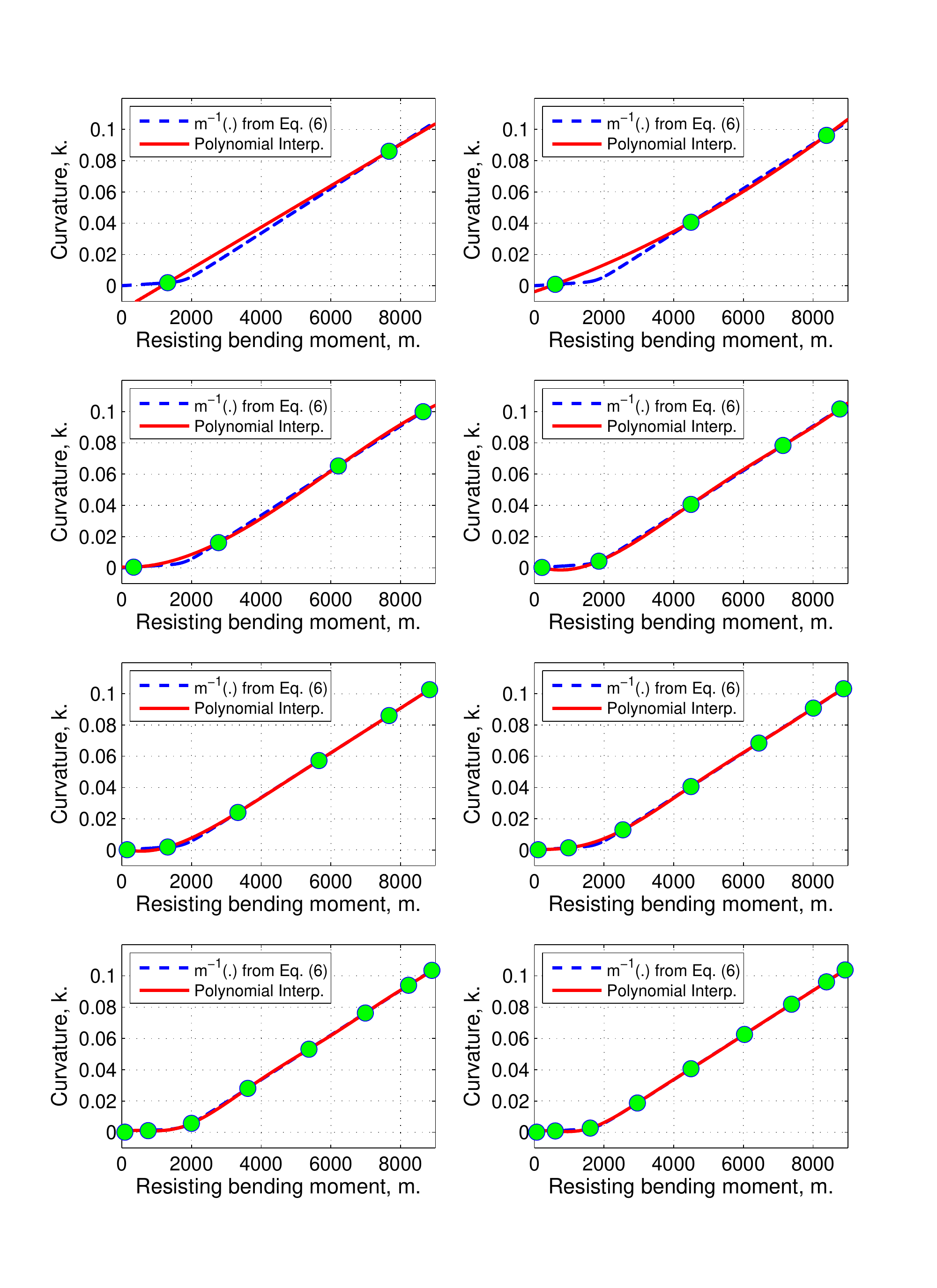}
\caption{The monomial form of polynomial interpolation of $m^{-1}(\cdot)$ is performed for the bilinear constitutive law in Eq. (37) using Eqs. (11-16). The green points illustrate the Chebyshev nodes defined by Eqs. (14) and (15), and a set of $N+1$ nodes means a polynomial of $N^{th}$ degree.}
\end{figure}

\begin{figure}[!t]
\centering
\includegraphics[scale=0.45]{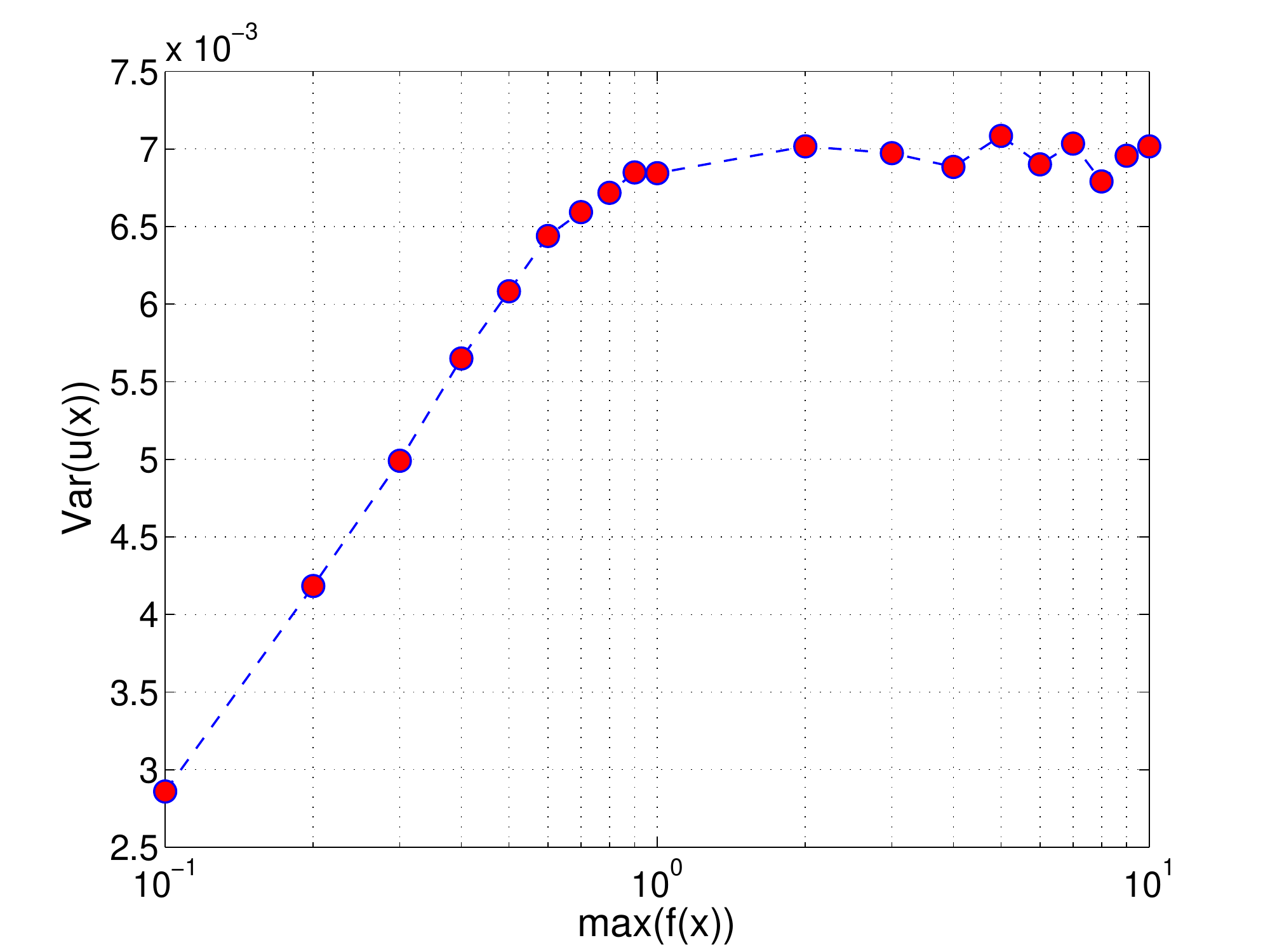}
\caption{The variance of the tip vertical displacement using 10,000 MC simulations for different truncation values of $f(x)$ having the PDF of LN, indicating the insignificance of the PDF tail's effect on the variance for truncation values larger than one.}
\end{figure}

Fig. (5) shows the components of the $\textbf{VRF}$ and $\textbf{S}(\kappa)-\delta(\kappa)\textbf{M}$ using a fourth degree polynomial interpolation of curvature-resisting bending moment. The response variance using the interpolation with different degrees of polynomial are shown in Fig. (6) in comparison with that of the MC simulations. Fig. (6-a) shows convergence for the interpolation-based approach as the polynomial degree increases. The converged variances derived analytically are very close to the variances determined through MC simulation, as illustrated in Fig. (6). While the MC simulation starts converging after about 1000 simulations, the method presented in this paper converges well after a polynomial of fifth to 10th degree. The relative error of analytical results with respect to the variance using 10,000 MC simulations are shown in Fig. (6-c) for different polynomial degrees.

Note that the responses generated in the MC simulation unlike Eq. (19) do not involve the polynomial interpolation of curvature-resisting bending moment and are calculated directly by
\begin{equation}
u(x)=
\int_0^x
k(s)\textup{sgn}(M(s))G(x,s)ds
\end{equation}
where curvatures are given by Eqs. (6-7) using linear interpolation within the pairs $( k,m( k))$. Such approach guarantees that the MC simulation, as the only verification benchmark, is not subject to the approximations of polynomial interpolation.

\begin{figure}[!t]
\centering
\includegraphics[scale=0.6]{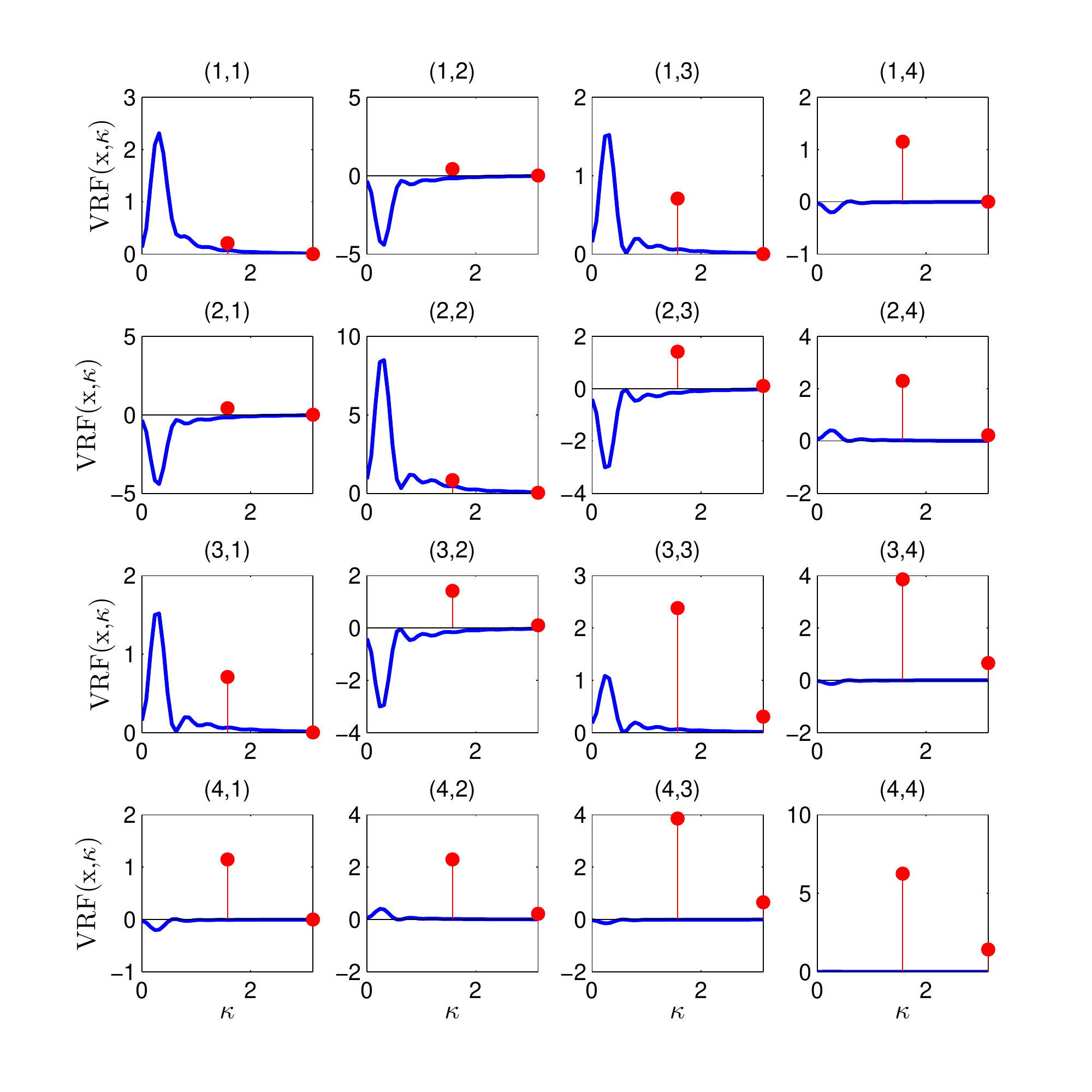}
\caption{The VRF components at $x=16$ using a fourth degree polynomial interpolation are plotted as blue lines. The terms of the components of $\textbf{S}(\kappa)-\delta(\kappa)\textbf{M}$ are represented for $\kappa_\delta=\pi /2$ by the red circles at $(n\kappa_\delta,a_{n}(i,j))$ and belong to $1+f(x)$ with $f(x)$ having the PDF of UN. The horizontal axes of the plots are the wave number ($\kappa$), and the plot titles $(i,j)$ indicate the component of the functions. The response variance is the sum of the blue curves' ordinates multiplied by $a_{n}(i,j)$ at $n\kappa_\delta$.}
\end{figure}

\begin{figure}[!t]
\centering
\includegraphics[scale=0.51]{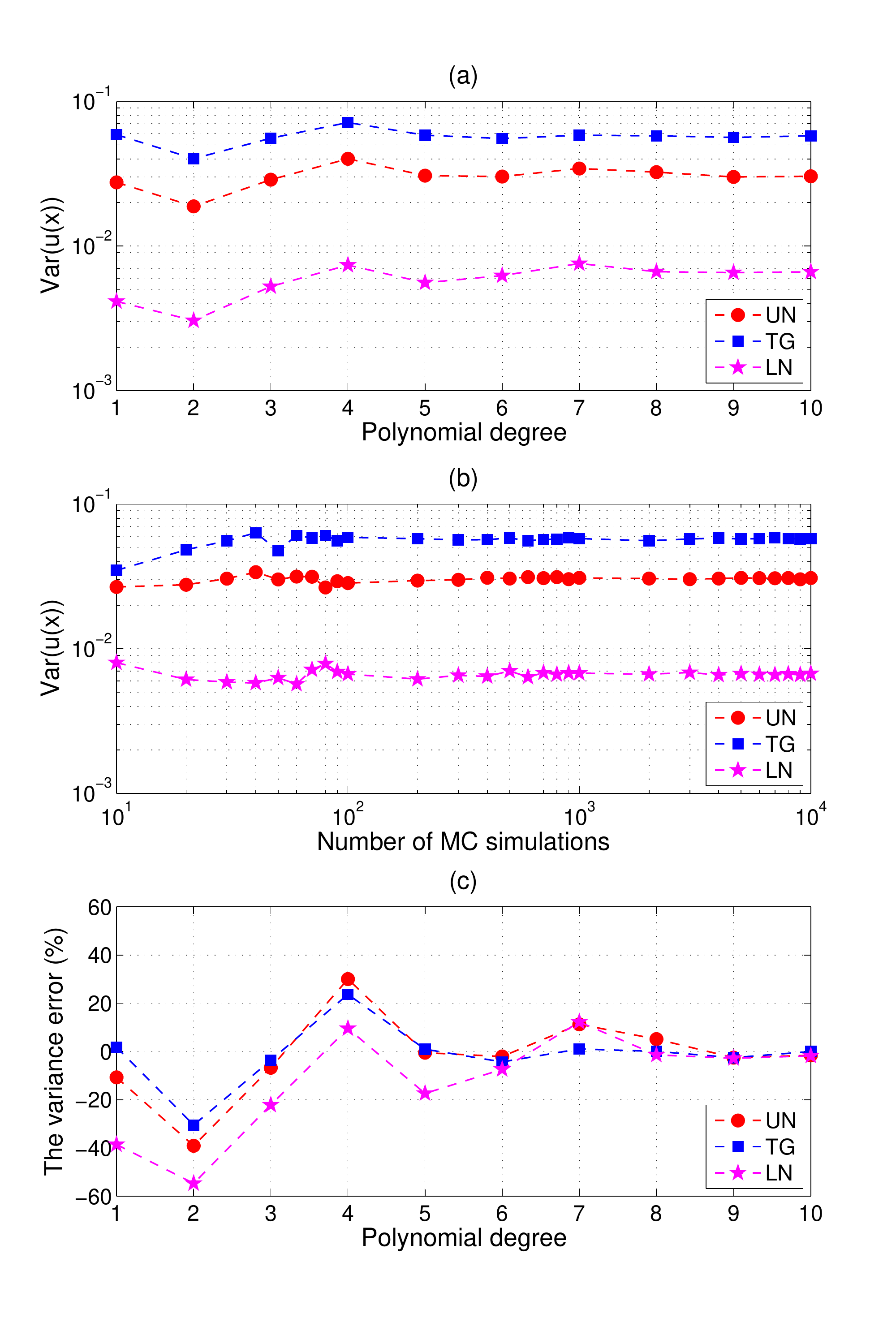}
\caption{The tip displacement variance for the bilinear constitutive law using (a) the monomial form of polynomial interpolation and (b) MC simulation, and (c)  the relative error of analytical method with respect to the results of 10,000 MC simulations. Note that calculations are based on $\kappa_\delta=\pi/2$.}
\end{figure}

\newpage
\clearpage
\section{Conclusion}
This paper generalizes the concept of the VRF to the response of stochastic statically determinate  Euler-Bernoulli beams having arbitrary functional forms of the constitutive law (i.e. Cauchy elastic materials). The new formulation is such that once the inverse of the nominal resisting bending moment with respect to the beam's curvature is interpolated by a polynomial function, the variance is determined by the inner product of a VRF matrix with a matrix containing the SDFs and statistical moments of the random field describing the resisting bending moment uncertainty. The interpolation-based approach certifies the closed-form VRFs already obtained for root constitutive laws and is tested to estimate the response variance of a cantilever having a bi-linear constitutive law by means of the VRF matrix. The accuracy of the VRFs is verified by the minor discrepancies among the predicted response variance values from the VRF with that obtained by MC simulation.

Another significance of this work is that the derivations presented in this paper open the possibility to compute VRFs for statically indeterminate structures having arbitrary Cauchy elastic constitutive laws using the Generalized Variability Response Function (GVRF) method. For statically indeterminate structures, the integrand in the expression for the response variance (e.g. Eq. (21)) cannot be separated into the product of a deterministic function (i.e. the VRF) and properties of the stochastic field (i.e. the SDF and higher order statistics). The GVRF method is a numerical technique to compute approximate VRFs and have been demonstrated on various statically indeterminate, linear structures \cite{Papadopoulos2006a, Papadopoulos2006, Papadopoulos2012, Papadopoulos2014, giovanis2015adaptive, papadopoulos2015transient}. 
For nonlinear constitutive laws, GVRFs can only be approximated if the specific higher order statistical moments and correlation functions that affect response variance for statically determinate structures, along with their relative contributions, are known \cite{Teferra2012}.

\end{document}